\documentclass[reprint,amsmath,amssymb,aps,pre]{revtex4-1}

\usepackage{graphicx}
\usepackage{dcolumn}
\usepackage{bm}

 \usepackage{color}

\usepackage{amssymb}
\usepackage{amsmath}
\usepackage{graphicx}
\usepackage{amsbsy}

\def\bal#1\eal{\begin{align}#1\end{align}}
\newcommand\beq{\begin{equation}}
\newcommand\eeq{\end{equation}}
\newcommand\beqa{\begin{eqnarray}}
\newcommand\eeqa{\end{eqnarray}}
\newcommand{\nn}{\nonumber\\}
\newcommand{\area}{S}

\newcommand{\xxi}{n}

\newcommand{\eeta}{\phi}

\newcommand{\ex}{\text{ex}}

\newcommand{\rcp}{\text{rcp}}

\newcommand{\eff}{\phi_{\text{eff}}}

\newcommand{\yeff}{y_{\text{eff}}}

\begin{document}



\title{Equation of state of polydisperse hard-disk mixtures in the high-density regime}


\author{Andr\'es Santos}
\email{andres@unex.es}
\homepage{https://www.eweb.unex.es/eweb/fisteor/andres/}
\author{Santos B. Yuste}
\email{santos@unex.es}
\homepage{https://www.eweb.unex.es/eweb/fisteor/santos/}
\author{Mariano L\'{o}pez de Haro}
\email{malopez@unam.mx}
\homepage{http://xml.cie.unam.mx/xml/tc/ft/mlh/}  \altaffiliation[ on sabbatical leave from ]{Instituto de Energ\'{\i}as Renovables, Universidad Nacional Aut\'onoma de M\'exico (U.N.A.M.),
Temixco, Morelos 62580, M{e}xico}
\affiliation{Departamento de F\'{\i}sica and Instituto de Computaci\'on Cient\'{\i}fica Avanzada (ICCAEx), Universidad de
Extremadura, E-06006 Badajoz, Spain}
\author{Vitaliy Ogarko}
\email{vitaliy.ogarko@uwa.edu.au}
\affiliation{University of Western Australia, Crawley WA 6009, Australia }
\date{\today}

\begin{abstract}
A proposal to link the equation of state of a monocomponent hard-disk fluid to the equation of state of a polydisperse hard-disk mixture is presented. Event-driven molecular dynamics simulations are performed to obtain data for the compressibility factor of the monocomponent fluid and of 26 polydisperse mixtures with different size distributions. Those data are used to assess the proposal and to infer the values of the compressibility factor of the monocomponent hard-disk fluid in the metastable region from those of mixtures in the high-density region. The collapse of the curves for the different mixtures is excellent in the stable region. In the metastable regime, except for two mixtures in which crystallization is present, the outcome of the approach exhibits a rather good performance. The simulation results indicate that a (reduced) variance of the size distribution larger than about $0.01$ is sufficient to avoid crystallization and explore the metastable fluid branch.
\end{abstract}



\maketitle
\section{Introduction}
\label{Intro}

The popularity in statistical physics of hard-core (hard-rod, hard-disk, hard-sphere, and hard-hypersphere) models for monocomponent fluids is undeniable. This is mostly due to the relative simplicity of their intermolecular interaction potentials. These models have also been important in the development of numerical simulation methods such as the Metropolis Monte Carlo algorithm \cite{MRRTT53} and the molecular dynamics (MD) method \cite{AW57}, which were first used in connection with monocomponent hard disks  and monocomponent hard spheres in a box, respectively. Nevertheless and despite this simplicity, except for the hard-rod case, no exact analytical expressions for the corresponding free energies of these systems are available. Therefore, most of their qualitative features, such as the existence of a stable fluid branch, a (freezing) fluid-solid phase transition at a given packing fraction $\phi_\text{f}$, a region of fluid-solid  coexistence, and a stable solid (crystalline) branch,  have been determined mostly from computer simulations. It is usual to present the equilibrium phase diagram for such fluids as a graph in the thermodynamic planes pressure $p$ versus density $\rho$ or compressibility factor $Z\equiv \beta{p}/{\rho}$ (where $\beta\equiv 1/k_BT$, with $k_B$ the Boltzmann constant and $T$ the absolute temperature) versus packing fraction $\phi\equiv v_d\rho\sigma^d$ [$\sigma$ being the diameter of the $d$-dimensional spheres, $d$ the dimensionality, and $v_d\equiv \left(\frac{\pi}{4}\right)^{d/2}\Gamma\left(1+\frac{d}{2}\right)$]. Note that $T$  enters in the description only as a scaling parameter, and so these systems are often referred to as athermal.

The case of the hard-disk (HD) system is especially interesting since it also presents a hexatic phase characterized by short-ranged positional order, but quasi-long-ranged orientational order. Further, for this two-dimensional hard-core system the nature of the freezing and melting transitions (first reported in the pioneering work of Alder \emph{et al.} \cite{AW62} and Alder, Hoover, and Wainwright \cite{AHW63}), the existence of a glass transition, and the location of the random close-packing fraction are still a source of debate. For more detailed information on these issues, see Refs.\ \onlinecite{B83,SR91,S99c,SK01,UST04,XBOH05,M06b,HHT06,DST06,DST07,KBR08,ZMSB10,PSK10,PZ10,TS10,BK11,XR11,HCFT12,SRSSTH14}.

In a different vein, polydispersity is known to be fundamental in studying important problems involving heterogeneous media, such as the mechanical properties of composite materials and the flow of fluids in porous media, and hence polydisperse systems also have received a lot of attention in the literature (see, for instance, Ref.\ \onlinecite{RS87} and references therein). Introducing polydispersity in size in hard-core models [in which case the packing fraction is $\phi=v_d \rho M_d$ with $M_q\equiv \int_0^\infty d\sigma\,\sigma^q f(\sigma)$ being the $q$th moment of the size distribution function $f(\sigma)$] is known to lead to a rich phenomenology (not present in the monocomponent models) that allows one, for instance, to avoid crystallization and modify the phase behavior \cite{D78,D80,D81,B00,PF04,TW12} or to deal in principle with real polydisperse systems as diverse as colloidal suspensions, granular matter, plastics, foams, powders, monolayers of mixtures adsorbed on a substrate, nanoparticles, micelles, food emulsions, and cell tissues (see Ref.\ \onlinecite{IT15} and the literature cited therein). Particular interest in polydisperse HD systems has focused on packing problems, random sequential absorption, and glassy behavior. Although the list is by no means exhaustive, the interested reader may refer to Refs.\ \onlinecite{HFJ86,BGOT86,BG89,TS89,TT91,MJ92,ASZW97,GKKB01,GB01,OO04,TKNV05,CC07,MSLB07,CCSB09,WH11,KT11,BMRRO12,MH12,H13,APF14,JLHMCT14,L14,K15,MSSWF16,ZM16,I16,IKCG16} for more information on various aspects of these problems, whose discussion lies beyond the scope of the present paper.

Apart from the other characteristic regions in the phase diagram that have been mentioned above, one important feature that all the monocomponent hard-core systems also present is that, beyond $\phi_\text{f}$, there is  a region of metastable fluid states that overlaps the fluid-solid coexistence region and also partly the crystalline branch. Accessing this metastable fluid branch is difficult using simulations. In particular, computing the values of the thermodynamic variables with enough accuracy in the metastable states is quite a challenge, so, in general, the metastable fluid branch remains as a largely unexplored ground. In previous work \cite{SYH99,S99d,HYS06,HYS08,SYH11,S12,S12c,SYHOO14}, we have provided different approximations for linking the equation of state (EoS) of a polydisperse  hard-core mixture and the EoS of the monocomponent system. Our approach has, for instance and among other things, allowed us \cite{SYH11,SYHOO14} to infer the EoS of the (three-dimensional) hard-sphere (HS) fluid in the metastable fluid region from high-density simulation data of polydisperse HS mixtures. It also led us to estimate and put some order in a wealth of values for the random close-packing fraction of polydisperse HS mixtures from the knowledge of the random close-packing fraction of the monocomponent system \cite{SYHOO14}.

The aim of this paper has several facets. On the one hand, given a polydisperse HD mixture with a certain size distribution at a packing fraction $\phi$, we will attempt to find an \emph{effective} monocomponent HD fluid such that the free energy and the EoS of the former system can be mapped onto those of the latter. On the other hand, simulation data for the compressibility factor of a large number of polydisperse HD mixtures with different size distributions will also be presented. These data will then be used to assess the mapping ``polydisperse mixture $\leftrightarrow$ monocomponent fluid.'' In particular, we will (a) check the collapse of all the mixture curves into a master one when plotted in the right variables and (b) use the high-density data of the mixtures to infer the EoS of a HD fluid in the metastable region. This latter aspect allows us to circumvent in part the difficulties associated with the simulation of the fluid-solid transition and the accessibility to the metastable fluid region in the monocomponent system. However, although for many years there has been interest in polydisperse systems, particularly bidisperse mixtures with particles of similar sizes for instance to determine whether and when those systems phase separate into  subsystems with different compositions, such problems and similar ones lie beyond the scope of this paper and will not be addressed here.

The paper is organized as follows. In Sec.\ \ref{sec2} we present the development linking the free energy and EoS of a polydisperse HD mixture to those of the monocomponent system.  Section \ref{sec3} presents the simulation results of the compressibility factor of a variety of polydisperse HD mixtures with different size distributions and the assessment of the mapping ``polydisperse mixture $\leftrightarrow$ monocomponent fluid,'' including the inference of the compressibility factor of the monocomponent HD fluid in the metastable region. The paper concludes in Sec.\ \ref{sec4} with further discussion and some concluding remarks.

\section{Mapping between the equation of state of the polydisperse mixture and that of the monocomponent system}
\label{sec2}

Recently, a fundamental-measure-theory (FMT) approach to link through an effective packing fraction $\eff$ the EoS of a monocomponent (three-dimensional) HS fluid to the EoS of a polydisperse HS mixture has been derived by application of some consistency conditions \cite{S12,S12c,SYHOO14,S16}. For completeness, the derivation is summarized in the Appendix. In this approach, the excess free energy per particle ($a^\ex$) of the mixture may be expressed in terms of the one of the monocomponent HS fluid ($a_s^\ex$) as
\beq
\beta a^\ex(\phi)+\ln(1-\phi)=\frac{\alpha}{\lambda}\left[\beta a_s^\ex(\eff)+\ln(1-\eff)\right],
\label{17}
\eeq
where the effective packing fraction $\phi_{\text{eff}}$ of the monocomponent  fluid   is  related to the packing fraction  $\phi$ of the polydisperse  mixture through
\begin{equation}
\frac{\phi_{\text{eff}}}{1-\phi_{\text{eff}}}=\frac{1}{\lambda}\frac{\phi}{1-\phi}.
\label{18a}
\end{equation}
Here,  $\lambda=m_3/m_2^2\geq 1$ and $\alpha=\lambda/m_2 \leq \lambda$ for three-dimensional systems,  $m_q\equiv M_q/M_1^q$ being the $q$th dimensionless moment.

In turn, taking into account the thermodynamic relation
\beq
\label{Zfroma}
Z(\phi)=1+\phi\frac{\partial \beta a^\ex(\phi)}{\partial \phi},
\eeq
the mapping between the compressibility factor of the monocomponent system ($Z_s$) and that of the polydisperse mixture ($Z$) that is then obtained from Eq.\ \eqref{17} may be expressed as
\beq
\phi Z(\phi)-\frac{\phi}{1-\phi}=\alpha\left[\eff Z_s(\eff)-\frac{\eff}{1-\eff}\right].
\label{19}
\eeq
Equivalently, the inverse of the thermodynamic relation \eqref{Zfroma}, namely,
\beq
\label{afromZ}
\beta a^\ex(\phi)=\int_0^\phi d\phi'\frac{Z(\phi')-1}{\phi'},
\eeq
allows one to recover Eq.\ \eqref{17} from Eq.\ \eqref{19}.

An interesting consequence of Eq.\ \eqref{19} is that one can invert it to \emph{infer} the monocomponent EoS from that of the polydisperse fluid. The degree of collapse of the mapping from different functions $Z(\phi)$ onto a \emph{common} function $Z_s(\phi_{\text{eff}})$ is an efficient way of assessing Eq.\ \eqref{19} without having to use an externally imposed EoS.

\begin{figure}[tbp]
 \hspace{-1cm}
\includegraphics[width=0.9\columnwidth]{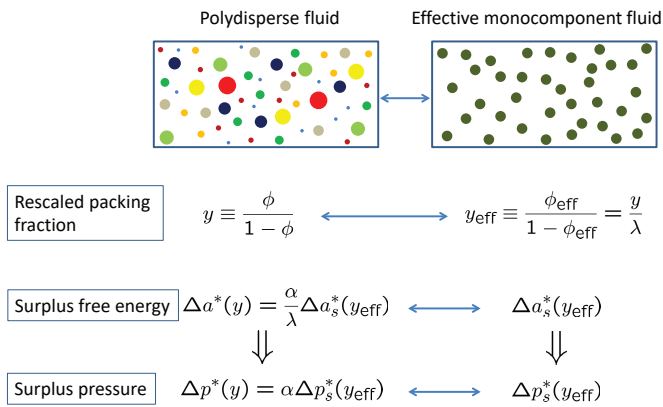}
\caption{Schematical view of the polydisperse $\leftrightarrow$ monocomponent mapping represented by Eqs.\ \eqref{17}--\eqref{19}.} \label{sketch}
\end{figure}

Equations \eqref{17}--\eqref{19} lend themselves to an insightful physical interpretation \cite{S16}. First, note that the ratio
\beq
\label{y}
y\equiv\frac{\phi}{1-\phi}
 \eeq
represents a \emph{rescaled} packing fraction, i.e., the ratio between the volume  occupied by the spheres and the remaining \emph{void} volume. Thus, Eq.\ \eqref{18a} dictates that the effective monocomponent fluid associated with a given mixture has a rescaled packing fraction $\yeff\equiv \eff/(1-\eff)$ that is $\lambda$ times smaller than that of the mixture. Next,
\beq
\label{Delta_p}
\Delta p^*(y)\equiv\phi Z(\phi)-\frac{\phi}{1-\phi}
\eeq
represents a (reduced) ``surplus'' pressure with respect to the ideal-gas value \emph{corrected} by the void volume.
Analogously, we can define the ``surplus''  free energy per particle
\beq
\Delta a^*(y)\equiv\beta a^\ex(\phi)+\ln(1-\phi)
\eeq
as the difference between the (reduced) free energy per particle  and the ideal-gas value corrected by the void volume. In terms of those quantities, Eqs.\  \eqref{17} and \eqref{19} establish that the surplus free energy $\Delta a^*$ and pressure $\Delta p^*$ of the polydisperse fluid are just proportional to their respective monocomponent counterparts $\Delta a^*_s$ and  $\Delta p^*_s$.
This is schematically depicted in Fig.\ \ref{sketch}.
While the surplus free energy of the polydisperse fluid is never larger than that of the effective monocomponent fluid (since $\alpha/\lambda\leq 1$),  the surplus pressure $\Delta p^*$ can be larger than, equal to, or smaller than $\Delta p^*_{s}$ (since $\alpha-1$ has not a definite sign).

It should be stressed that the proposal implied by Eq.\ \eqref{19} may be interpreted in two directions. On the one hand, if $Z_s$ is known as a function of $\eff$, then one can readily compute $Z$ as a function of $\phi$ [$\eff$ and $\phi$ being of course related through Eq.\ \eqref{18a}]. On the other hand, if $Z(\phi)$ is known in a high enough density region, then the values of $Z_s(\eff)$ may be inferred from those of the mixture even in regions where obtaining them from simulation is either difficult or not feasible, such as the metastable fluid branch.

Although initially derived for three-dimensional HS systems \cite{S12c,SYHOO14}, the physical interpretation of Eqs.\ \eqref{17}--\eqref{19} suggests their heuristic generalization to any dimensionality $d\neq 3$. In that case, the parameters $\lambda$ and $\alpha$  can be determined by imposing consistency with the second and third virial coefficients of the mixture \cite{S16}. This leads to
\begin{equation}
\lambda=\frac{\bar{B}_2-1}{b_2-1}\frac{b_3-2b_2+1}{\bar{B}_3-2\bar{B}_2+1},\quad
\alpha=\lambda^2\frac{\bar{B}_2-1}{b_2-1},
\label{26}
\end{equation}
where $\bar{B}_n\equiv B_n/(v_d M_d)^{n-1}$ and $b_n\equiv B_n/(v_d \sigma^d)^{n-1}$ are reduced virial coefficients of the mixture and the monocomponent fluid, respectively ($B_n$ being the standard virial coefficients).
Of course, $\lambda=m_3/m_2^2\geq 1$ and $\alpha=\lambda/m_2 \leq \lambda$ are again recovered from Eq.\ \eqref{26} in the three-dimensional case.

In what follows, we will particularize to HD systems ($d=2$), in which case $v_2=\frac{\pi}{4}$, $b_2=2$,  $b_3=4(4/3-\sqrt{3}/\pi)\simeq 3.12802$, and the second virial coefficient of the polydisperse HD mixture is given exactly by
\beq
\bar{B}_2=1+m_2^{-1}.
\label{B2b}
\eeq
On the other hand, given a size distribution $f(\sigma)$, the \emph{exact} third virial coefficient of a polydisperse mixture of \emph{additive} HDs is not expressible in terms of moments. Its explicit expression is \cite{S16}
\begin{align}
B_3=&\frac{\pi}{2}\int_0^\infty d\sigma_1 \,f(\sigma_1)\int_0^\infty d\sigma_2\, f(\sigma_2)\sigma_{12}^2\nonumber\\
&\times\int_0^\infty d\sigma_3\, f(\sigma_3)\area_{\sigma_{13},\sigma_{23}}(\sigma_{12}),
\label{23b}
\end{align}
where $\sigma_{ij}=\frac{1}{2}(\sigma_i+\sigma_j)$ and
\bal
\area_{a,b}(r)=&a^2\cos^{-1}\frac{r^2+a^2-b^2}{2ar}
+b^2\cos^{-1}\frac{r^2+b^2-a^2}{2br}\nn
&-\frac{1}{2}\sqrt{2r^2(a^2+b^2)-(b^2-a^2)^2-r^4}
\label{27}
\eal
is the intersection area of two circles of radii $a$ and $b$ whose centers are separated by a distance $r$.

Therefore, for HD fluids Eqs.\ \eqref{17} and \eqref{19} become
\beq
\beta a^\ex(\phi)+\ln(1-\phi)=\frac{\lambda}{m_2}\left[\beta a_s^\ex(\eff)+\ln(1-\eff)\right],
\label{17_2D}
\eeq
\beq
\phi Z(\phi)-\frac{\phi}{1-\phi}=\frac{\lambda^2}{m_2}\left[\eff Z_s(\eff)-\frac{\eff}{1-\eff}\right],
\label{19_2D}
\eeq
respectively, where now
\begin{equation}
\lambda=\frac{b_3-3}{(\bar{B}_3-1)m_2-2}.
\label{26_2D}
\end{equation}
Notice that Eq.\ \eqref{19_2D} is equivalent to
\beq
\Delta p^*(y)=\frac{\lambda^2}{m_2}\Delta p^*_s(y/\lambda).
\label{19_2Db}
\eeq

It is worthwhile noting that in the Scaled Particle Theory (SPT) for HD systems \cite{RFL59,HFL61,LHP65} one simply has $\Delta p_s^*(y)=y^2$ and $\Delta p^*(y)=m_2^{-1} y^2$, so that Eq.\ \eqref{19_2Db} is identically satisfied for \emph{arbitrary} values of $\lambda$.
In the more general case, however, the  polydisperse $\leftrightarrow$ monocomponent  mapping depends on $\lambda$, as given by Eq.\ \eqref{26_2D}, to guarantee that the third virial coefficient of the mixture is exactly retained. On the other hand, this latter requirement implies an added difficulty since, as said before, the exact determination of $\bar{B}_3$ via Eq.\ \eqref{23b} cannot be carried out from the knowledge of just the first few moments of the size distribution $f(\sigma)$. Therefore, it seems practical to replace the exact formula \eqref{23b} by an approximate simpler one. In Ref.\ \cite{SHY05} three of us proposed an approximate expression of $B_3$ for nonadditive $d$-dimensional HS mixtures. Its particularization to additive HD mixtures yields \cite{S16}
\beq
\bar{B}_3^{\text{app}}=1+\frac{b_3-1}{m_2}.
\label{B3bapp}
\eeq
Using this approximation in Eq.\ \eqref{26_2D}, we obtain $\lambda=1$ and $\alpha=m_2^{-1}$. Thus, $\eff=\phi$ and Eqs.\ \eqref{17_2D}, \eqref{19_2D}, and \eqref{19_2Db} reduce to
\beq
\beta a^\ex(\phi)+\ln(1-\phi)=\frac{1}{m_2}\left[\beta a_s^\ex(\phi)+\ln(1-\phi)\right],
\eeq
\beq
Z(\phi)-\frac{1}{1-\phi}=\frac{1}{m_2}\left[Z_s(\phi)-\frac{1}{1-\phi}\right],
\label{Zd=2}
\eeq
\beq
\Delta p^*(y)=\frac{1}{m_2}\Delta p^*_s(y).
\eeq
Interestingly enough, this mapping was  derived earlier from a different method \cite{SYH99,HYS08}.

Equation (\ref{19_2D}) [together with Eqs.\ (\ref{18a}), (\ref{23b}), (\ref{27}), and \eqref{26_2D}], on the one hand, and Eq.\ (\ref{Zd=2}), on the other hand,  provide two different (approximate) connections
between the compressibility factor of polydisperse HD mixtures of given size distribution and that of the monocomponent HD fluid.

For later comparison, we also include here another theoretical mapping proposed by Barrio and Solana \cite{BS99,BS00b,BS06}, which reads
\bal
Z(\phi)-1=&\left[\frac{1+m_2^{-1}}{2}-\left(b_3 \frac{1+m_2^{-1}}{4}-\frac{\bar{B}_3}{2}\right)\phi\right]\nn
&\times\left[Z_s(\phi)-1\right].
\label{BS}
\eal
Again the mapping in Eq.\ \eqref{BS} is consistent with $\bar{B}_2$ and $\bar{B}_3$ but cannot be expressed in terms of moments unless $\bar{B}_3$ is replaced by $\bar{B}_3^{\text{app}}$.

The three above proposals for the mapping between $Z$ and $Z_s$ will be assessed in Sec.\ \ref{sec3}.

\section{Results}
\label{sec3}

\subsection{Systems examined}
In order to test the usefulness of the approximations implied by Eqs.\ \eqref{19_2D}, \eqref{Zd=2}, and \eqref{BS}, the following  classes of size distributions have been chosen. First,  binary (B) mixtures having a discrete composition characterized by
\beq
f(\sigma) = (1 - x)\delta(\sigma - a) + x \delta(\sigma - a w),
\eeq
where $a$ is the small diameter, $w$ is the ratio of the big to the small diameter, and $x$ is the mole fraction of the big species. The associated $q$th-order moment is $M_q=a^q(1-x+x w^q)$.
Next, the top-hat (TH) distribution
\beq
f(\sigma) = \frac{1}{a(w - 1)}\Theta(\sigma - a)	\Theta(a w - \sigma),
\eeq
where $\Theta(x)$ is the Heaviside step function. In this case, $M_q=a^q(w^{q+1}-1)/(w-1)(q+1)$.
Finally, we consider
\beq
f(\sigma) = \frac{aw\sigma^{-2}}{w - 1}\Theta(\sigma - a)\Theta(a w - \sigma),
\eeq
so that the distribution decays in the interval $a < \sigma < a w$ as an inverse power (IP) law of second order. Here, $M_1=aw(\ln w)/(w-1)$ and $M_q=a^q (w^q-w)/(w-1)(q-1)$ for $q\neq 1$.

\begin{table}
   \caption{Values of $\bar{B}_2$, $\bar{B}_3$, $\bar{B}_3^{\text{app}}$, $\lambda$, and $\alpha$ for the different polydisperse HD mixtures examined in this work.}\label{table1}
\begin{ruledtabular}
\begin{tabular}{lccccccc}
Label&$x$&$w$&$\bar{B}_2$&$\bar{B}_3$&$\bar{B}_3^{\text{app}}$&$\lambda$&$\alpha$\\
\hline
B1&$0.5$&$1.4$ &$1.97297$&$3.06851$&$3.07050$&$1.01630$&$1.00495$\\
B2&$0.3$&$2$ &$1.88947$&$2.88496$&$2.89282$&$1.07412$&$1.02621$\\
B3&$0.25$&$4$&$1.64474$&$2.35456$&$2.37201$&$1.26816$&$1.03689$\\
B4&$0.14$&$6$&$1.48983$&$2.02135$&$2.04237$&$1.50420$&$1.10830$\\
B5&$0.07$&$14$&$1.24902$&$1.51371$&$1.52991$&$2.03367$&$1.02989$\\
B6&$0.04$&$22$&$1.16661$&$1.34172$&$1.35456$&$2.51284$&$1.05206$\\
B7&$0.03$&$30$&$1.12502$&$1.25584$&$1.26605$&$2.76326$&$0.95463$\\
B8&$0.025$&$40$&$1.09520$&$1.19471$&$1.20258$&$2.82185$&$0.75802$\\
TH1&&$1.1$&$1.99924$&$ 3.12635$&$ 3.12641$&$1.00046$&$1.00016$ \\
TH2 &&$1.2$&$1.99725$&$ 3.12196$&$ 3.12217$&$1.00166$&$1.00056$\\
 TH3&&$1.4$&$1.99083$&$ 3.10780$&$ 3.10849$&$1.00553$&$1.00181$\\
  TH4&&$2$&$1.96429$&$ 3.04943$&$ 3.05202$&$1.02136$&$ 1.00593$\\
 TH5&&$3$&$1.92308$&$ 2.95922$&$ 2.96432$&$1.04515$&$ 1.00831$\\
 TH6&&$5$&$1.87097$&$ 2.84588$&$ 2.85343$&$1.07264$&$ 1.00210$\\
 TH7&&$10$&$1.81757$&$ 2.73069$&$ 2.73980$&$1.09530$&$ 0.98083$\\
 TH8&&$20$&$1.78563$&$ 2.66232$&$ 2.67183$&$1.10447$&$ 0.95835$\\
 TH9&&$30$&$1.77417$&$ 2.63789$&$ 2.64744$&$1.10663$&$ 0.94808$\\
 TH10&&$100$&$1.75743$&$ 2.60233$&$ 2.61181$&$1.10845$&$ 0.93062$\\
 TH11&&$500$&$1.75150$&$ 2.58977$&$ 2.59920$&$1.10865$&$ 0.92367$\\
 IP1 &&$1.4$&$1.99062$&$ 3.10734$&$ 3.10805$&$1.00570$&$ 1.00195$\\
IP2 &&$1.6$&$1.98179$&$ 3.08789$&$ 3.08927$&$1.01114$&$ 1.00378$\\
 IP3&&$1.8$&$1.97170$&$ 3.06566$&$ 3.06779$&$1.01744$&$ 1.00588$\\
 IP4&&$2$&$1.96091$&$ 3.04191$&$ 3.04483$&$1.02428$&$ 1.00813$\\
 IP5&&$2.4$&$1.93850$&$ 2.99266$&$ 2.99715$&$1.03883$&$ 1.01280$\\
 IP6&&$3$&$1.90521$&$ 2.91960$&$ 2.92631$&$1.06138$&$ 1.01975$\\
 IP7&&$4$&$1.85414$&$ 2.80783$&$ 2.81762$&$1.09835$&$ 1.03041$\\
   \end{tabular}
 \end{ruledtabular}
 \end{table}

Table \ref{table1} contains a list of the 26 mixtures (B1--B8, TH1--TH11, IP1--IP7) examined in this paper, together with the corresponding values of their exact second  and third  virial coefficients, as given by Eqs.\ \eqref{B2b} and \eqref{23b}, respectively. Note that, except in the binary cases, the values of $\bar{B}_3$ need to be obtained numerically. Table \ref{table1} also includes the values of $\lambda$, as obtained from Eq.\ \eqref{26_2D}, and of $\alpha=\lambda^2/m_2$, as well as those of the approximate third virial coefficient, Eq.\ \eqref{B3bapp}.

A convenient measure of the ``degree of dispersity'' in a mixture can be taken as the reduced variance  $m_2-1$ of the size distribution. Equivalently, according to Eq.\ \eqref{B2b}, the degree of dispersity can be measured by $2-\bar{B}_2=1-m_2^{-1}$. In a B mixture at a fixed value of $w$, $m_2$ takes a maximum value $m_2=(1+w)^2/4w$ at $x=1/(1+w)$; this maximum value monotonically increases (almost linearly) without upper bound as $w$ increases. In the case of IP mixtures, one has $m_2=(w-1)^2/w (\ln w)^2$, which again grows unbounded (but much more slowly) with increasing $w$. On the other hand, $m_2=4(1+w+w^2)/3(1+w)^2$ for TH mixtures, this quantity being now upper bounded by $m_2=\frac{4}{3}$.
Taking all of this into account, we can order the 26 mixtures of Table \ref{table1} in ascending degree of dispersity as
TH1--TH3, IP1, IP2, B1, IP3, TH4, IP4, IP5, TH5, IP6, B2, TH6, IP7, TH7--TH11, B3--B8.
Interestingly, even though $\bar{B}_3$ is not expressible in terms of moments, it turns out that the same ordering is obtained if the degree of dispersity is measured by $b_3-\bar{B}_3$. The same happens if the dispersity criterion is $\lambda-1$, except for the permutations B2$\leftrightarrow$TH6 and IP7$\leftrightarrow$TH7.

\begin{figure}[tbp]
 \hspace{-1cm}
\includegraphics[width=0.9\columnwidth]{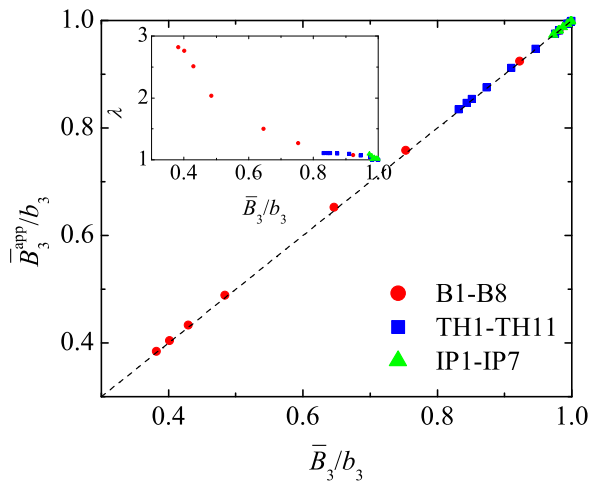}
\caption{Scatter plot of $\bar{B}_3^{\text{app}}/b_3$ vs $\bar{B}_3/b_3$ for the $26$ mixtures considered in Table \ref{table1}. The straight dashed line represents the equality $\bar{B}_3^{\text{app}}/b_3=\bar{B}_3/b_3$. The inset shows a scatter plot of $\lambda$ vs $\bar{B}_3/b_3$.} \label{fig:B3_scatt}
\end{figure}

It can be observed from Table \ref{table1} that $\bar{B}^{\text{app}}_3$ tends to overestimate $\bar{B}_3$ but otherwise it is always a very good approximation, with a maximum relative deviation of $1.07$\%  (mixture B5). This is graphically illustrated in Fig.\ \ref{fig:B3_scatt}, which shows that
$\bar{B}_3^{\text{app}}\simeq \bar{B}_3$, even when the third virial coefficient is more than 40\% smaller than its monocomponent value.

Table \ref{table1} also shows that $\lambda\leq 1.11$, except in the mixtures B3--B8, where $\lambda$ increases rather rapidly with dispersity (see inset in Fig.\ \ref{fig:B3_scatt}). It seems paradoxical that $\lambda$ can be as large as $\lambda=2.82$ (mixture B8) even though $\delta \bar{B}_3\equiv \bar{B}_3^{\text{app}}-\bar{B}_3$ is relatively small (for instance, $\delta \bar{B}_3=0.008$ in the case of B8). In order to understand this, note that Eq.\ \eqref{26_2D} can be recast as $\lambda^{-1}=1-\delta \bar{B}_3/(b_3-3)(\bar{B}_2-1)$. As long as $\delta \bar{B}_3/(b_3-3)\ll \bar{B}_2-1$, as happens typically if $\bar{B}_2>1.7$, $\lambda$ is only slightly larger than $1$. On the other hand, as dispersity increases, $\bar{B}_2$ becomes closer to $1$ and, therefore, $\lambda$ visibly departs from $1$.

\subsection{Molecular dynamics simulations}

We have used event-driven MD simulations with a modification of the Lubachevsky--Stillinger algorithm \cite{LS90,DTS05,OL12,OL13} to compute the compressibility factors of  polydisperse HD mixtures having the 26 size distributions described by Table \ref{table1}, as well as for  the monocomponent HD system. Starting from zero packing fraction,  the
system is compressed by allowing the diameter of the disks to grow linearly in time with a dimensionless rate $\Gamma$, while the kinetic energy $\mathcal{E}$ is kept constant using a rescaling thermostat procedure. Therefore, in each case, a single MD run was employed to span a wide range of packing fractions (from $\phi\approx 0$ to $\phi\approx 0.85$).

A two-dimensional square unit cell of area $\mathcal{A}$ containing $\mathcal{N}$ particles was considered and, as usual,  periodic boundary conditions  were used to mimic an infinite system.
Given the velocities ($\mathbf{v}_i$ and $\mathbf{v}_j$) just before contact  of two interacting particles $i$ and $j$ (with masses $\mu_i$ and $\mu_j$, respectively) and given their relative position ($\mathbf{r}_{ij}=\mathbf{r}_{i}-\mathbf{r}_{j}$), the velocities ($\mathbf{v}_i'$ and $\mathbf{v}_j'$)
after the collision are derived from conservation of linear momentum as $\mu_{i,j}\mathbf{v}_{i,j}'= \mu_{i,j}\mathbf{v}_{i,j}\pm \Delta \mathbf{p}$. Since we are dealing with smooth (frictionless) particles, only the normal component
of the change of momentum, $\Delta \mathbf{p}^{(n)}=(\Delta \mathbf{p}\cdot \widehat{\mathbf{n}})\widehat{\mathbf{n}}$ (where  $\widehat{\mathbf{n}} = \mathbf{r}_{ij}/{r}_{ij}$ is the unit vector in the normal direction), is affected during the collision. It is obtained from
\beq
\Delta \mathbf{p}^{(n)} = -2\mu_{ij}\mathbf{v}_c^{(n)},
\eeq
with $\mu_{ij}=\mu_i\mu_j/(\mu_i+\mu_j)$ being the reduced mass and $\mathbf{v}_c^{(n)}$ being the normal component of the relative velocity of the
contact-point of the particles. The latter is calculated taking into account the expanding disk diameters
(for growing particles) as
\beq
\mathbf{v}_c^{(n)}=\left[(\mathbf{v}_i-\mathbf{v}_j)\cdot \widehat{\mathbf{n}}- \frac{1}{2}\left(\dot{\sigma}_i+\dot{\sigma}_j\right)\right]\widehat{\mathbf{n}}.
\eeq

In our simulations, the diameter $\sigma_i(t)$ of particle $i$ was grown according to the linear law $\dot{\sigma_i}=\Gamma \sqrt{\mathcal{E}/\mathcal{M}} \sigma_i(t)/\sigma_{\text{max}}(t)$, where $\mathcal{E}$ and $\mathcal{M}$ are the total energy and mass, respectively, $\sigma_{\text{max}}(t)$ is the largest diameter in the system at time $t$, $\Gamma$ is the (constant) dimensionless growth rate, and the ratio $\sigma_i(t)/\sigma_{\text{max}}(t)$ is independent of time.  In this convention, we have $\sigma_{\text{max}}(t)= \Gamma \sqrt{\mathcal{E}/\mathcal{M}} t$ and $\sigma_{\text{min}}(t)= \Gamma \sqrt{\mathcal{E}/\mathcal{M}} w^{- 1} t$, where $\sigma_{\text{min}}(t)$ is the diameter of the smallest particle and $w=\sigma_{\text{max}}(t)/\sigma_{\text{min}}(t)$ is constant, as desired. This ensures that the size distribution during the process is maintained and thus disk areas relative to the mean are constant over time, but the mean disk area increases uniformly with time. According to this protocol, the packing fraction grows quadratically in time as $\phi(t)=\frac{\pi}{4}\sum_i \sigma_i^2(t)/\mathcal{A}=\frac{\pi}{4}(\mathcal{E}\mathcal{N}/\mathcal{M}\mathcal{A})(M_2/\sigma_{\text{max}}^2)\Gamma^2 t^2$.

The compressibility factor $Z$  at temperature $k_B T = \mathcal{E}/{\mathcal{N}}$ is calculated from
the total exchanged momentum in all interparticle collisions during a short time period $\Delta t$, namely,
\beq
Z = 1+\sum_{\text{coll}} \frac{|\Delta \mathbf{p}_{ij}|r_{ij}}{2 \mathcal{E} \Delta t},
\eeq
where  ${r}_{ij}$ accounts for the distance over which momentum  is transmitted. Note that, over time, the additional energy created during collisions would accelerate the particles, but this is avoided by a periodic rescaling of the average particle velocity to hold the mean temperature constant. In fact, the time period $\Delta t$ (typically $400$ events per particle) was chosen so that the total change in the kinetic energy due to growth stays below $1\%$.
\begin{figure}[tbp]
 \hspace{-1cm}
\includegraphics[width=0.9\columnwidth]{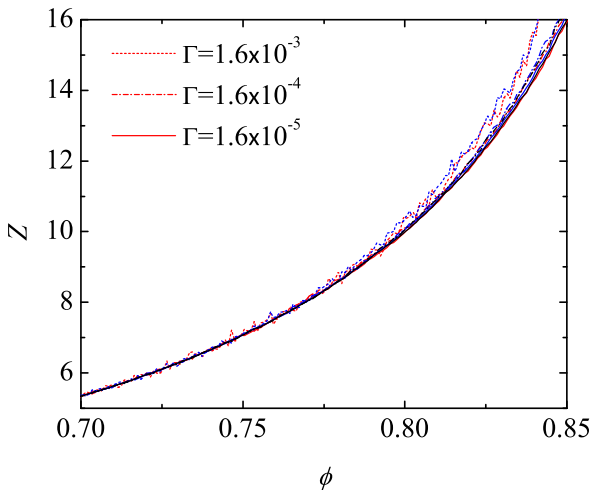}
\caption{Plot of $Z(\phi)$ as obtained from our simulations for the  HD mixture B5. The curves correspond to $\mathcal{N}=2^{12}$ (red), $\mathcal{N}=2^{13}$ (blue), and $\mathcal{N}=2^{14}$ (black), with growth rates $\Gamma=1.6\times 10^{-3}$ (dotted), $\Gamma=1.6\times 10^{-4}$ (dash-dotted), and $\Gamma=1.6\times 10^{-5}$ (solid). Note that the case $\mathcal{N}=2^{14}$ with $\Gamma=1.6\times 10^{-3}$ has not been included.} \label{fig:B5}
\end{figure}

If the growing is sufficiently slow, the system will approximately be in  equilibrium during the densification process and one can rather efficiently gather quasi-equilibrium data as a function of density. We used slow growth rates $\Gamma=1.6 \times 10^{-4}$ to $1.6 \times 10^{-5}$ and a number of particles $\mathcal{N}=2^{12}=4096$, except in the most disparate mixtures (B4--B8), where $\mathcal{N}=2^{14}=16\,384$.

One may reasonably wonder whether the choices of growth rate and  number of particles do have an important influence on our simulation results. As an illustration of the effect of varying both $\mathcal{N}$ and $\Gamma$ on the robustness of our calculations, in Fig.\ \ref{fig:B5} we show results for mixture B5. Although not shown, the ones for other mixtures display similar behavior. It is clear  that, for a given number of particles, the rate $\Gamma=1.6\times 10^{-3}$ is too fast for our purposes if $\phi\gtrsim 0.77$, but the rates $\Gamma=1.6\times 10^{-4}$ and $\Gamma=1.6\times 10^{-5}$ lead to practically  indistinguishable results in the region of interest $\phi\leq 0.85$. On the other hand, for a fixed growth rate,  the results for $\mathcal{N}=2^{12}$, $\mathcal{N}=2^{13}$, and $\mathcal{N}=2^{14}$ are almost identical.  In general, then, an excellent selection seems to be $\mathcal{N}=2^{14}$  and $\Gamma=1.6\times 10^{-5}$. This guarantees meaningful data in the density range of interest. It is clear  also that even the choice $\mathcal{N}=2^{12}$ and $\Gamma=1.6\times 10^{-4}$ would represent a reasonable compromise between robustness and computation time. Nevertheless, in the extremely dense region $\phi>0.85$, not considered in this paper, increasingly slower growth rates would be necessary to guarantee a good equilibration.

\begin{figure}[tbp]
\includegraphics[width=0.91\columnwidth]{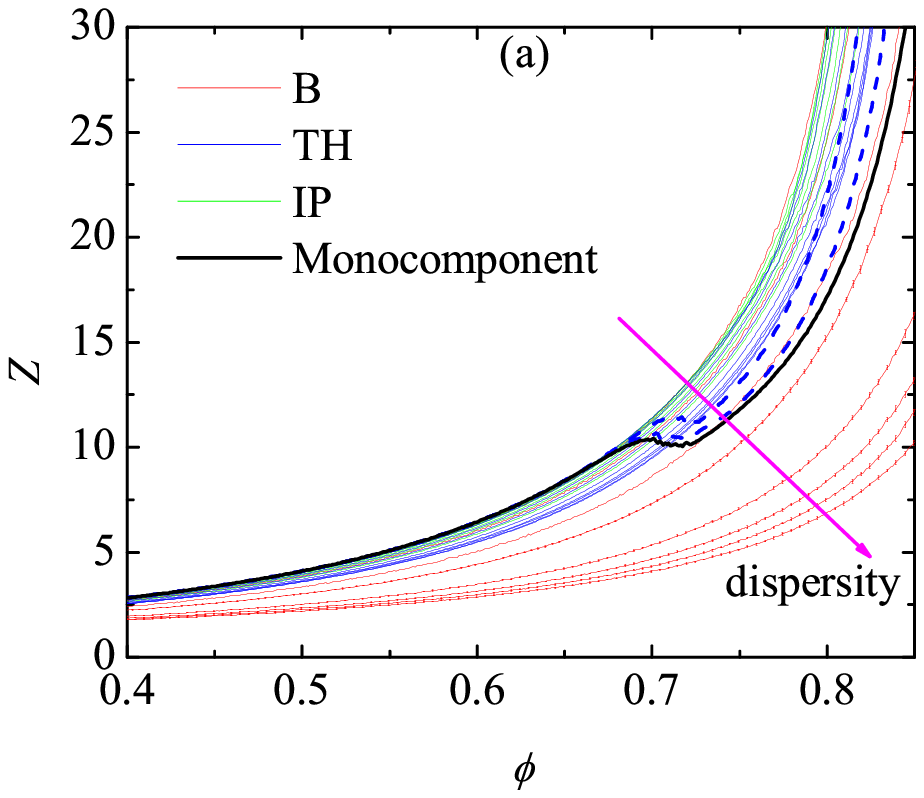}\\
\includegraphics[width=0.91\columnwidth]{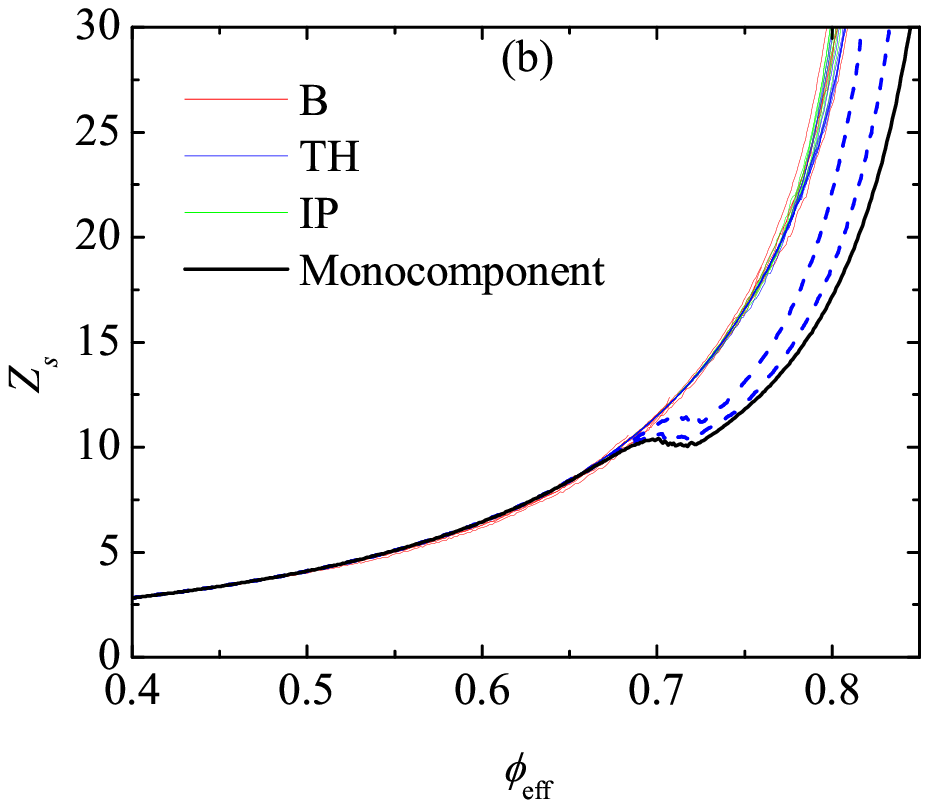}
\caption{(a) Plot of $Z(\phi)$ as obtained from our simulations for the monocomponent HD fluid and for various polydisperse HD mixtures. (b) Plot of the inferred monocomponent compressibility factor $Z_s(\eff)$ as obtained from Eqs.\ \protect\eqref{18a} and \protect\eqref{19_2D}, complemented by Eqs.\ \protect\eqref{23b} and \protect\eqref{26_2D}. The thick black solid line corresponds to the monocomponent system, while the thin red (B1--B8), blue (TH3--TH11), and green (IP1--IP7) solid lines correspond to polydisperse systems. The two thick blue dashed lines refer to mixtures TH1 and TH2. Note the tiny error bars in the curves B4--B8 of panel (a).} \label{fig1}
\end{figure}

Thanks to the large number (${\sim}{10^4}$) of pairs $\{\phi(t),Z(t)\}$ obtained as direct output for a given size distribution, we could afford to smooth out the data with a bin size $\delta\phi= 10^{-3}$. This in turn allowed us to estimate the error bars from the fluctuations of $Z$ within each bin. The error bar of $Z$ was observed to generally increase with $\phi$, but it typically remained close to $1\%$ at $\phi\simeq 0.85$ [see Fig.\ \ref{fig1}(a)].

\begin{figure}[tbp]
\includegraphics[width=0.91\columnwidth]{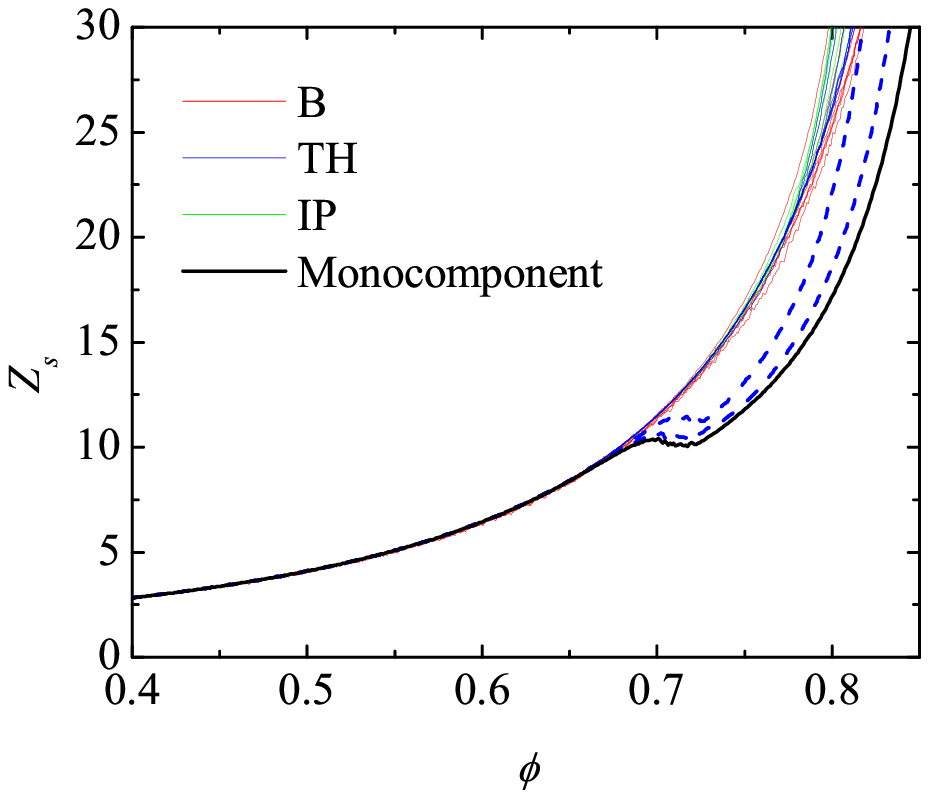}
\caption{Plot of the inferred monocomponent compressibility factor $Z_s(\phi)$ as obtained from Eq.\ \protect\eqref{Zd=2}. The meaning of the curves is the same as in Fig.\ \protect\ref{fig1}.} \label{fig2}
\end{figure}

\begin{figure}[tbp]
\includegraphics[width=0.91\columnwidth]{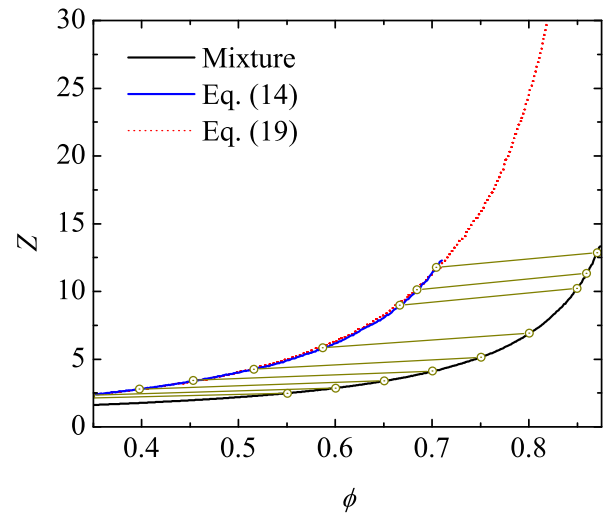}
\caption{Plot of the compressibility factor for the mixture B8 (black solid line) and for the inferred monocomponent compressibility factor  as obtained from Eq.\ \protect\eqref{19_2D} (blue solid line) and from Eq.\ \protect\eqref{Zd=2} (dotted red line). The circles joined by straight lines denote pairs $(\eff,\phi)$ as given by Eq.\ \protect\eqref{18a}.} \label{fig4}
\end{figure}

\begin{figure}[tbp]
\includegraphics[width=0.91\columnwidth]{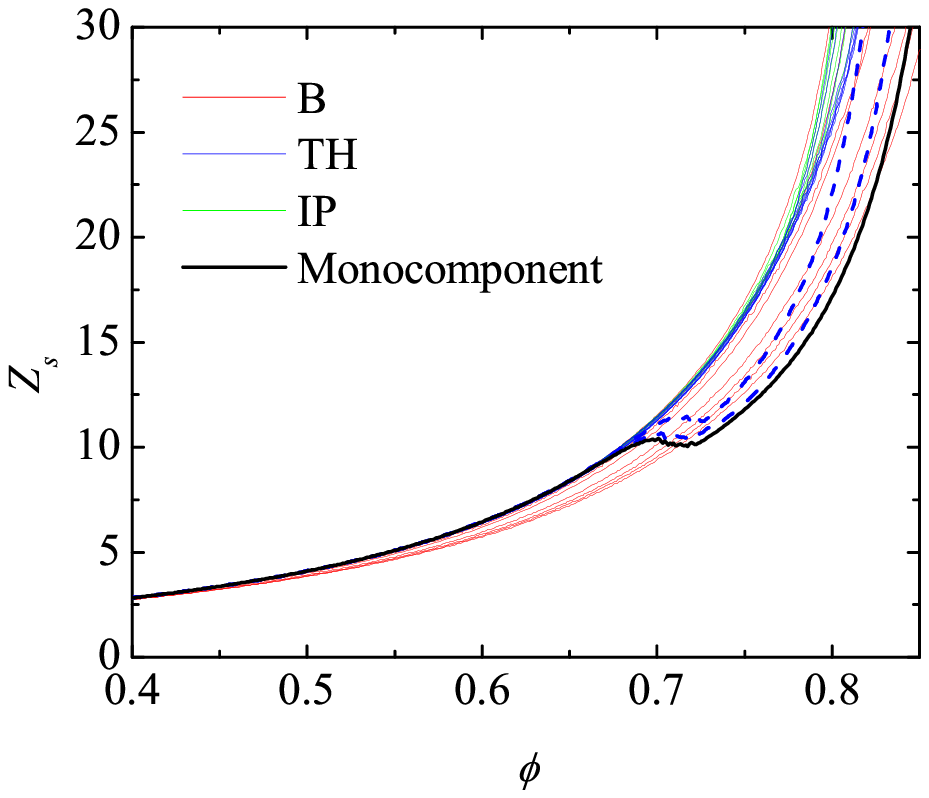}
\caption{Plot of the inferred monocomponent compressibility factor $Z_s(\phi)$ as obtained from Eq.\ \protect\eqref{BS} with  $\bar{B}_3$  given by Eq.\ \protect\eqref{23b}. The meaning of the curves is the same as in Fig.\ \protect\ref{fig1}.} \label{fig3}
\end{figure}

\subsection{Assessment of the mapping}
Figure \ref{fig1}(a) shows all the functions $Z(\phi)$  for those 26 mixtures and for the monocomponent system. As expected, each mixture differs in the values of $Z$ for a common $\phi$. If the mapping ``polydisperse mixture $\leftrightarrow$ monocomponent fluid''  as given by Eq.\ \eqref{19_2D} works, a high degree of collapse of all the curves should be expected when the inferred monocomponent quantity $Z_s(\eff)$  is plotted instead of $Z(\phi)$. This is shown in Fig.\ \ref{fig1}(b), where  an excellent collapse of all the mixture curves is observed in the stable region ($\phi_{\text{eff}}\lesssim 0.7$), the collapse keeping being rather good (although, of course, not perfect) in the metastable region ($\phi_{\text{eff}}> 0.7$), except  for the mixtures TH1 and TH2, in which the ratio between the diameters of the biggest and the smallest disk is less than or equal to $w=1.2$. Those latter curves exhibit crystallization effects in that region and thus follow trends similar to that of the monocomponent fluid. This observation agrees with the result of Speedy \cite{S99c}, who also pointed out that, in the case of equimolar binary HD mixtures, freezing to a mixed crystal occurs when $w < 1.2$ while they may reach the metastable fluid region when $w \geq 1.3$. On the other hand, a higher degree of dispersity (say, $m_2\gtrsim 1.01$) allows one to frustrate equilibration to a crystal phase and explore the metastable fluid branch, which is practically inaccessible for the monocomponent fluid.
Interestingly, we have checked (not shown) that the extrapolations in the metastable domain of known accurate EoS for the monocomponent HD fluid \cite{H75,SHY95,L01b,LS01,LS04} agree fairly well with the inferred values of $Z_s$ in that domain.

Figure \ref{fig2} shows the inferred $Z_s(\phi)$ as obtained with the simplified mapping  given by Eq.\ \eqref{Zd=2}, i.e., with $\lambda\to 1$ or $\bar{B}_3\to\bar{B}_3^{\text{app}}$. Despite this simplification,  we get a collapse of the mixture curves practically as good as the one shown in Fig.\ \ref{fig1}(b), although the mapping corresponding to Eq.\ \eqref{19_2D} is somewhat more accurate in the metastable region. Notwithstanding this, it should be pointed out that the mapping \eqref{19_2D} has two drawbacks with respect to the mapping \eqref{Zd=2}. First, it requires the numerical evaluation of the scaling parameter $\lambda$ via Eqs.\ \eqref{23b} and \eqref{26_2D}. Second,  since $\phi_{\text{eff}}<\phi$, one has to go to higher packing fractions of the mixture to get access to the high-density region of the monocomponent system.  This is a drawback because the simulation results for the polydisperse mixtures become noisier, and hence more unreliable, as the packing fraction of the mixture increases, due to diverging collision rates. The previous feature is illustrated in Fig.\ \ref{fig4}, where we have joined with  straight lines the points connecting the packing fraction $\phi$ of the mixture B8 with the corresponding effective value $\eff$ of the monocomponent fluid. Thus, according to Eq.\ \eqref{19_2D}, the values of $Z$ for the mixture B8 on the region $0<\phi<0.87$ map onto values of $Z_s$ in the narrower region $0<\eff<0.70$. In contrast, Eq.\ \eqref{Zd=2} allows one to infer $Z_s$ at the same packing fractions as for the mixture (i.e., the corresponding points would be joined by vertical lines, not shown).

As said at the end of Sec.\ \ref{sec2}, a mapping alternative to Eqs.\ \eqref{19_2D} and \eqref{Zd=2} is represented by Eq.\ \eqref{BS}. The associated inferred $Z_s(\phi)$ is shown in Fig.\ \ref{fig3}. Here, the true values of $\bar{B}_3$ have been used, but we have checked that the curves are practically indistinguishable from those obtained if $\bar{B}_3$ is replaced by $\bar{B}_3^{\text{app}}$ as given in Eq.\ \eqref{B3bapp}.
While the mapping \eqref{BS} does a good job for not too disparate mixtures (say, $m_2<2$ or, equivalently, $\bar{B}_2>1.5$), it clearly fails for the mixtures B4--B8, even in the stable region. In fact, Eq.\ \eqref{BS} is inconsistent with the exact limit of a binary mixture in which the small disks are point particles \cite{HYS02}.

\subsection{Mapping the monocomponent system onto itself?}
As said in Sec.\ \ref{sec2}, Eq.\ \eqref{19_2D} becomes Eq.\ \eqref{Zd=2} if $\bar{B}_3\to B_3^\text{{app}}$, i.e., $\lambda\to 1$. However, for very disparate mixtures (e.g., mixture B8), the scaling parameter $\lambda$ can clearly deviate from $\lambda=1$, as can be seen from Table \ref{table1} and the inset of Fig.\ \ref{fig:B3_scatt}. Despite this, we have seen from Figs.\ \ref{fig1}--\ref{fig4} that both approximations \eqref{19_2D} and \eqref{Zd=2} yield practically the same inferred monocomponent compressibility factor $Z_s(\phi)$, even from the mixture B8. This suggests the possibility of combining Eqs.\ \eqref{19_2D} and \eqref{Zd=2} to map $Z_s(\phi)$ onto itself, as explained below.

Suppose we use as input a known monocomponent compressibility factor $Z_s^\text{in}$. Then, Eq.\ \eqref{19_2D} allows us to obtain the compressibility factor $Z$ for a given polydisperse fluid: $Z_s^\text{in}\to Z$. Next, insertion of that function $Z$ into Eq.\ \eqref{Zd=2} provides an output monocomponent function $Z_s^\text{out}$: $Z\to Z_s^\text{out}$. Proceeding in this way, we can define a mapping $Z_s^\text{in}\to Z_s^\text{out}$, which reads
\beq
\label{Z_s_map}
Z_s^\text{out}(\phi)=\frac{\lambda^2}{\lambda-(\lambda-1)\phi}Z_s^\text{in}\left(\frac{\phi}{\lambda-(\lambda-1)\phi}\right)-\frac{\lambda-1}{\phi}.
\eeq
This is a nontrivial monocomponent $\to$ monocomponent  nonlocal mapping. The degree of consistency $Z_s^\text{out}(\phi)\simeq Z_s^\text{in}(\phi)$ for $\lambda>1$ is an indirect measure of how equivalent the mappings \eqref{19_2D} and \eqref{Zd=2} are.
In terms of the scaled packing fraction \eqref{y} and the surplus pressure \eqref{Delta_p}, Eq.\ \eqref{Z_s_map} can be simply rewritten as
\beq
\label{Delta_ps}
\Delta p_s^{*\text{out}}(y)=\lambda^2 \Delta p_s^{*\text{in}}\left(\lambda^{-1}y\right).
\eeq
The consistency condition $\Delta p_s^{*\text{out}}(y)=\Delta p_s^{*\text{in}}(y)$ yields the self-similarity relation $\Delta p_s^{*}(y)=\lambda^2 \Delta p_s^{*}\left(\lambda^{-1}y\right)$, whose \emph{unique} solution is the SPT EoS $\Delta p_s^{*\text{SPT}}(y)=y^2$ or, equivalently, $Z_s^\text{SPT}(\phi)=1/(1-\phi)^2$.
Now, the interesting question is, How good is the mapping $Z_s^\text{in}(\phi)\to Z_s^\text{out}(\phi)\simeq Z_s^\text{in}(\phi)$, even though $Z_s(\phi)\neq Z_s^\text{SPT}(\phi)$?

\begin{figure}[tbp]
\includegraphics[width=0.91\columnwidth]{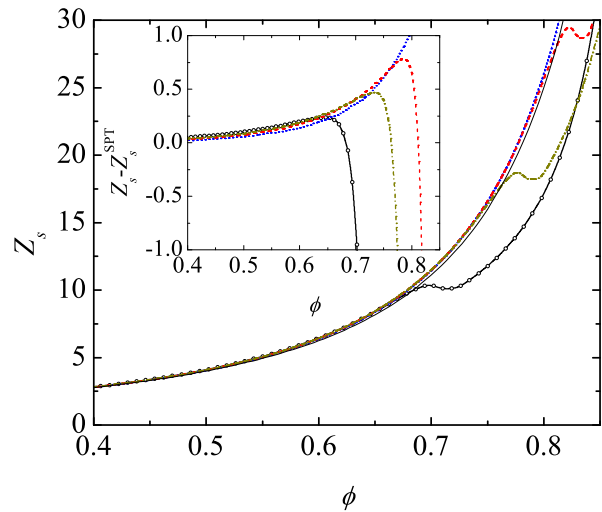}
\caption{Plot of our MD results for the monocomponent compressibility factor $Z_s^\text{in}(\phi)$ (---$\circ$---, black) and the output functions $Z_s^\text{out}(\phi)$ obtained from $Z_s^\text{in}(\phi)$ by using Eq.\ \eqref{Z_s_map} with $\lambda=1.5$ (--$\cdot$--$\cdot$--, dark yellow), $\lambda=2$ (-- -- --, red), and $\lambda=3$ ($\cdots$, blue). The (black) thin solid line represents the function $Z_s^\text{SPT}(\phi)$. The inset shows the differences $Z_s(\phi)-Z_s^\text{SPT}(\phi)$.} \label{fig:Zs_map}
\end{figure}

To address this question, Fig.\ \ref{fig:Zs_map} plots
our MD results for the monocomponent compressibility factor, $Z_s^\text{in}(\phi)$, together with the output functions $Z_s^\text{out}(\phi)$ obtained from it via Eq.\ \eqref{Z_s_map} with $\lambda=1.5$ (similar to the value corresponding to mixture B4), $\lambda=2$ (similar to the value corresponding to mixture B5), and $\lambda=3$ (larger than the value corresponding to mixture B8). As a reference, the SPT function $Z_s^\text{SPT}(\phi)$ is also plotted, while the deviations $Z_s(\phi)-Z_s^\text{SPT}(\phi)$ are displayed in the inset of Fig.\ \ref{fig:Zs_map}. We can observe that, as is well known, the SPT EoS slightly underestimates the compressibility factor. As was already seen in Figs.\ \eqref{fig1}, \eqref{fig2}, and \eqref{fig3}, the ``true'' monocomponent function $Z_s^\text{in}(\phi)$ exhibits a freezing transition at $\phi\approx 0.7$. On the other hand, part of the stable liquid branch of $Z_s^\text{in}(\phi)$ for $\phi<0.7$ can be  mapped onto $Z_s^\text{out}(\phi)$ for  a certain domain of the metastable region $\phi>0.7$. The larger the parameter $\lambda$, the larger the explored metastable region.

Of course, since $Z_s(\phi)\neq Z_s^\text{SPT}(\phi)$, the mapping $Z_s^\text{in}(\phi)\to Z_s^\text{out}(\phi)\simeq Z_s^\text{in}(\phi)$ is imperfect and depends on the value of $\lambda$. Nevertheless, both the main panel and the inset of Fig.\ \ref{fig:Zs_map} show a high degree of overlap of $Z_s^\text{out}(\phi)$ in the stable and metastable regions, even for relatively large values of $\lambda$.
Therefore, it can be tentatively conjectured that it might be possible to use Eq.\ \eqref{Z_s_map} to map the monocomponent EoS onto itself, extending it into the metastable region and thus obviating the need for polydispersity.

\section{Concluding Remarks}
\label{sec4}

In this paper we have presented a heuristic ansatz for the relationship between the excess Helmholtz free energy of a polydisperse HD mixture, $a^\ex$, and the one of the monocomponent HD fluid, $a_s^\ex$ [see Eq.\ \eqref{17_2D}]. Such an ansatz maintains the form introduced earlier for the three-dimensional case \cite{SYHOO14} and allowed us to derive a link between both compressibility factors in which $Z$ and $Z_s$ are evaluated at \emph{different} packing fractions [see Eq.\ \eqref{19_2D}].

The mapping may be used in the two directions: from a known compressibility factor $Z_s$ as a function of the packing fraction $\eff$ one may obtain the compressibility factor of \emph{any} polydisperse mixture at the corresponding packing fraction $\phi$; on the other hand, from values of $Z(\phi)$ at a high enough density one may compute the corresponding values of $Z_s(\eff)$ in difficult or inaccessible density regions from the simulation point of view. The only information required in the mapping are the second and third virial coefficients of the polydisperse mixture and those of the monocomponent system. Thus, the main asset of our approach is its relative simplicity.

In the two-dimensional case examined here,  the expressions of the required  virial  coefficients  of the mixture for a given size distribution are exactly known. It turns out that, in analogy with what happens in three dimensions,  the second virial coefficient is a rather simple function of the first two moments of the size distribution. On the other hand, in contrast to the case of three-dimensional HSs, the third virial coefficient  for a polydisperse HD mixture may not be expressed in terms of the moments of the size distribution, but is nevertheless amenable to explicit evaluation. If one makes a further approximation, namely, the replacement of the exact third virial coefficient by the approximate value $\bar{B}_3^{\text{app}}$ [also given in terms of the first two moments; see Eq.\ \eqref{B3bapp}], one gets another mapping [see Eq.\ \eqref{Zd=2}] which coincides with one derived earlier with a different method in which both compressibility factors are evaluated at the same packing fraction  \cite{SYH99,HYS08}. With these two proposals, and making use of the compressibility factors obtained from MD simulations for a variety of polydisperse HD mixtures that cover a wide density range, we have been able to derive the values of the compressibility factor of a monocomponent HD fluid in the metastable fluid branch beyond the fluid-solid phase transition.

While the collapse  of the curves corresponding to the very different mixtures, as obtained from the mapping \eqref{19_2D}, is not perfect [see Fig.\ \ref{fig1}(b)], it is quite reasonable both in the stable and metastable regions and allows one to identify systems in which some degree of crystallization may still be present. In this respect, although at this stage the issue remains as a conjecture, our results suggest that for a given polydisperse mixture, irrespective of its size distribution,  a value of the reduced second moment $m_2$  greater than $1.01$ is enough to frustrate equilibration to a crystal phase. This provides a general criterion to classify the systems that allow the meaningful inference of the EoS of the monocomponent HD, including the metastable fluid branch, from the knowledge of the high-density data of such polydisperse HD mixtures.

It also turns out that the inferred  EoS of the monocomponent system with the mapping \eqref{Zd=2} based on $\bar{B}_3^{\text{app}}$  is almost as accurate as the one without such an approximation (see Fig.\ \ref{fig2}) and superior to the (also simple) proposal of Barrio and Solana \cite{BS99,BS00b,BS06}, especially in the case of very disparate mixtures. Therefore, Eq.\ \eqref{Zd=2} seems to be a reasonable compromise between accuracy and simplicity.

As done  in the case of HSs in Ref.\ \cite{SYHOO14}, it would be tempting to try to
estimate the jamming packing fraction $\phi_J$ of a given mixture from the knowledge of the random close-packing fraction $\phi_\rcp$ of the monocomponent system. In fact, the reasonably good degree of collapse observed in Fig.\ \ref{fig1}(b) for very high packing fractions provides some support to the use of Eq.\ \eqref{18a} for such an estimate. Hence, in this case
\beq
\phi_J\approx \frac{\lambda}{\lambda+\phi_\rcp^{-1}-1},
\eeq
with $\lambda$ given by Eqs.\ \eqref{23b}, \eqref{27}, and \eqref{26_2D}.

Finally, while it could be argued that dealing with the thermodynamic properties of two-dimensional HDs may appear to be a merely academic problem, it has practical relevance to confined or adsorbed colloidal systems. In fact, for this latter case some interesting experiments have recently been published in which a canted colloidal monolayer in sedimentation equilibrium is used \cite{TAAD17}. These experiments confirm the existence of a first-order liquid-hexatic phase transition followed by an apparently second-order (or at least continuous) freezing transition. The assessment of possible artifacts due to size polydispersity, which arises both intrinsically and from out-of-plane fluctuations in the sedimentation equilibrium, is also addressed. Since sedimentation is a direct probe of compressibility, it is conceivable that the analytical approach presented in this paper could hopefully have direct application to this problem.

\begin{acknowledgments}
We want to thank Adri\'an Huerta for a fruitful exchange of correspondence. We are also grateful to the anonymous referees for suggesting the plots of Figs.\ \ref{fig:B3_scatt} and \ref{fig:Zs_map}. A.S., S.B.Y., and M.L.H. acknowledge the financial support of the Ministerio de Econom\'ia y Competitividad (Spain) through Grant No.\ FIS2016-76359-P and  the Junta de Extremadura (Spain) through Grant No.\ GR15104, both partially financed by ``Fondo Europeo de Desarrollo
Regional'' funds. M.L.H. also acknowledges the Consejo Nacional de Ciencia y Tecnolog\'ia (CONACYT, Mexico) for a sabbatical Grant.
\end{acknowledgments}

\appendix*

\section{Fundamental-measure-theory derivation of Eq.\ \eqref{17}}

In  FMT, the bulk excess free-energy density $\Phi$ of a three-dimensional HS fluid mixture is assumed to depend on the partial number densities only through the four SPT variables $\{n_\alpha\}$ \cite{RELK02}, the construction of a specific functional $\Phi(\{n_\alpha\})$ being guided by internal consistency conditions. Two basic conditions are (i) in the special case where one of the components is made of point particles $\Phi$ must decompose into an ideal-gas-like term  plus a term associated with the remaining components, and (ii) a common pressure must be obtained from the standard thermodynamic relation and from the reversible work needed to create a cavity  accommodating a particle of infinite diameter. Here we summarize the proof \cite{S12,S12c,S16} that the functionals satisfying (i) and (ii)  must necessarily have the form
\beq
\label{PhiPhi}
\Phi(\{n_\alpha\})=-n_0\ln(1-n_3)+\Psi(y)\frac{n_1n_2}{1-n_3},
\eeq
where
\beq
\label{yy}
y\equiv \frac{n_2^2}{12\pi n_1 (1-n_3)}
\eeq
is a scaled variable and  $\Psi(y)$  is an arbitrary dimensionless scaling function which can be determined from the free-energy density of the monocomponent system.

\subsection{Hard-sphere mixture}
Let us consider an {additive} HS mixture characterized by an arbitrary number of components with diameters $\{\sigma_i; i=1,2,\ldots\}$ and partial number densities $\{\rho_i=N_i/V; i=1,2,\ldots\}$, $N_i$ being the number of particles of species $i$ and $V$ being the volume. Other related quantities are the total number density $\rho=\sum_i\rho_i=N/V$, the mole fractions $\{x_i=\rho_i/\rho; i=1,2,\ldots\}$,   and the FMT bulk densities
\beq
    \xxi_0=\rho,
\quad
\xxi_1=\frac{1}{2}\rho M_1,
\quad
\xxi_2=\pi \rho M_2,
\quad
\xxi_3=\frac{\pi}{6}\rho M_3,
\eeq
where  $M_q=\sum_i x_i \sigma_i^q$ are size distribution moments. In particular, $\eeta\equiv\xxi_3$ is the packing fraction.

The pressure and chemical potentials are given by appropriate derivatives of the {excess} free energy (in units of $k_BT$) per unit volume $\Phi(\{\rho_i\})=\Phi(\rho;\{x_i\})$,
\begin{subequations}
\beq
    \beta p=\rho+\rho^2\left(\frac{\partial}{\partial\rho}\frac{\Phi}{\rho}\right)_{\{x_i\}},
    \label{A1}
    \eeq
    \beq
    \beta\mu_i^\ex=\left(\frac{\partial \Phi}{\partial\rho}\right)_{\{\rho_{j\neq i}\}}.
    \label{A1b}
    \eeq
\end{subequations}

\subsection{Physical conditions on $\Phi$}
Let us assume that  we add $N_0=\rho_0 V$ particles of zero diameter ($\sigma_0=0$). In such a case,  it can be proved that
\cite{S12}
  \beq
{ \lim_{\sigma_0\to 0}\Phi(\{\rho_0,\rho_1,\rho_2\ldots\})=-\rho_0\ln(1-\xxi_3)+\Phi(\{\rho_1,\rho_2,\ldots\})}.
 \label{A2}
  \eeq
This is the point-particle limit condition on the free energy density $\Phi$.

Another independent condition is related to the fact that, apart from the thermodynamic relation \eqref{A1}, the pressure can alternatively be obtained taking into account the reversible work needed to create a cavity  accommodating a particle of infinite diameter \cite{RELK02}:
\beq
{\beta p=\lim_{\sigma_i\to\infty}\frac{\beta\mu_i^\ex}{{\pi}\sigma_i^3/{6}}}.
\label{A3}
\eeq

\subsection{Truncatable free energies and FMT}

A free energy density $\Phi(\{\rho_i\})$ is said to be \emph{truncatable} if it depends on the series of densities $\{\rho_i\}$ and the series of diameters $\{\sigma_i\}$ only through the packing fraction $\eeta$ and a \emph{finite} number of moments \cite{S02}.
In the special case of a FMT \cite{R89,TCM08}, the number of moments is reduced to three:
\beq
\Phi(\{\rho_i\})\to \Phi(n_0,n_1,n_2,n_3).
\eeq
Application of Eq.\ \eqref{A2} implies \cite{S12}
   \beq
  {{\Phi(\{n_\alpha\})=-n_0\ln(1-n_3)+n_1 n_2 H(n_3,y)}},
  \label{A5}
  \eeq
where $y$ is defined by Eq.\ \eqref{yy} and the dimensionless function $H(n_3,y)$ remains so far undetermined.
On the other hand, imposing consistency between Eqs.\ \eqref{A1} and \eqref{A3}, and using Eq.\ \eqref{A1b}, we obtain
  \beq
{{(1-\xxi_3)\frac{\partial\Phi}{\partial \xxi_3}=\xxi_0-\Phi+\sum_{\alpha=0}^2\xxi_\alpha\frac{\partial\Phi}{\partial \xxi_\alpha}}}.
\label{A6}
\eeq

The constraints  \eqref{A5} and \eqref{A6} can be combined to yield  Eq.\ \eqref{PhiPhi} \cite{S12c,S16}, which can be rewritten as
\bal
\Phi(\{\xxi_\alpha\})=&-\xxi_0\ln(1-\xxi_3)+\frac{4\pi\xxi_1^2}{\xxi_2}\Big[\beta a_s^\ex\left(\frac{y}{1+y}\right)\nn
&-\ln(1+y)\Big],
\label{A7}
\eal
where
$a_s^\ex(\eeta)$ is the (bulk) excess free energy {per particle} of the {monocomponent} system.
Thus, the thermodynamic properties of a HS mixture of total packing fraction $\eeta$ are expressed in terms of those of a pure HS fluid with an \emph{effective} packing fraction
\beq
\eeta_{\text{eff}}=\frac{y}{1+y}=\left[1+\frac{1-\eeta}{\eeta}\frac{M_1M_3}{M_2^2}\right]^{-1}\leq \eeta.
\eeq
Extension to the inhomogeneous case is achieved by appropriate replacements of the variables $\{n_\alpha\}$ by the FMT  weighted densities \cite{HMOR15}.

In the homogeneous case, it is straightforward to obtain the excess free energy per particle $\beta a^\ex=\Phi/n_0$ from Eq.\ \eqref{A7}. The result is given by Eqs.\ \eqref{17} and \eqref{18a} with
\beq
\alpha=\frac{M_1^3M_3}{M_2^3},\quad \lambda=\frac{M_1M_3}{M_2^2}.
\eeq

\bibliography{D:/Dropbox/Mis_Dropcumentos/bib_files/Liquid}

\end{document}